\documentclass[journal]{IEEEtran}
\usepackage{amsmath,amsfonts}
\usepackage{algorithmic}
\usepackage{array}
\usepackage[caption=false,font=normalsize,labelfont=sf,textfont=sf]{subfig}
\usepackage{textcomp}
\usepackage{cite} 
\usepackage{stfloats}
\usepackage{url}
\usepackage{verbatim}
\usepackage{comment}
\usepackage{graphicx}
\usepackage{siunitx}
\usepackage{array}
\usepackage[table]{xcolor}
\def\BibTeX{{\rm B\kern-.05em{\sc i\kern-.025em b}\kern-.08em
    T\kern-.1667em\lower.7ex\hbox{E}\kern-.125emX}}
\usepackage{balance}
\begin{document}
\title{Novel 1-bit Hybrid Reconfigurable Intelligent Surface}
\author{Sajedeh Keshmiri, Graduate Student Member, IEEE, Suren Jayasuriya, Senior Member, IEEE, and Mohammadreza F. Imani, Member, IEEE
\thanks{The authors are with the School of Electrical, Computer, and Energy Engineering, Arizona State University, Tempe, AZ 85287 USA (e-mail: skeshmi1@asu.edu; sjayasur@asu.edu; mohammadreza.imani@asu.edu).}}


\maketitle

\begin{abstract}
This paper proposes a novel 1-bit hybrid reconfigurable intelligent surface (HRIS) designed to sense the incident signal's angle of arrival (AoA) and redirect it toward desired directions. This device consists of independently tunable resonant patch elements loaded with PIN diodes. To introduce sensing capabilities, a portion of the signal incident on each element is coupled to a parallel plate waveguide (PPWG) through small rectangular slots. Two coaxial connectors are then used to collect the signal coupled to the PPWG. We use compressive sensing techniques and a multi-layer perceptron to analyze this signal and detect AoA. Further, pre-coded phase randomization is implemented by varying slot sizes to suppress undesired quantization lobes. The proposed HRIS is simple, low-cost, and can pave the way for intelligent wireless communication, power transfer, and sensing without needing feedback loops.
\end{abstract}


\section{Introduction}
\IEEEPARstart{R}{econfigurable} Intelligent Surfaces (RISs) consist of tunable elements that can dynamically redirect electromagnetic waves to control the propagation environment. They can optimize signal paths, mitigate interference, and enhance network security \cite{yu2011light,kaina2014shaping,renzo2019smart,basar2019wireless,wang2023applications,elmossallamy2020reconfigurable,bae2024overview,zhang2024secure,alamzadeh2021reconfigurable,pei2021ris,trichopoulos2022design,gros2021reconfigurable,popov2021experimental,shi2024ultra}. To effectively engineer the propagation environment, RISs require information about desired directions. However, this requirement is often overlooked in the current literature on RIS-empowered wireless networks, which typically assumes that the full or partial channel state information is readily available. Instead, the focus has been on designing reflecting metasurface configurations that can achieve the desired reconfigurable reflection patterns, or investigate their capabilities in improving wireless links\cite{trichopoulos2022design,gros2021reconfigurable,popov2021experimental,shi2024ultra,diaz2017generalized}.

Over the years, various methods for obtaining channel state at the RIS have been proposed. The primary method is to measure the BS-RIS-UE (Base Station-RIS-User Equipment) channel and use that to decompose the separate channels of BS-RIS and RIS-UE\cite{zhou2020joint,mutlu2024joint, wei2022joint,swindlehurst2022channel}. This method is computationally expensive and susceptible to noise. Another method involves space-time coding metasurfaces, but this adds complexity due to the need for temporal modulation \cite{fang2022accurate}. A different strategy dedicates certain RIS elements exclusively for channel sensing, which, while effective, results in lower spatial resolution and discontinuous phase reflective phase profiles\cite{ma2020smart,taha2021enabling}. A more recent approach is hybrid RISs (HRISs)\cite{alamzadeh2021reconfigurable,alexandropoulos2023hybrid, ghazalian2023joint,zhang2023channel} where sensing capabilities are integrated into RISs configuration by modifying all or a few of the constituent meta-atoms to couple a portion of the incident signal into a waveguide for sampling and analysis. Using the collected signal, the angle of arrival (AoA) or channel state information can be retrieved, promising a way toward a fully autonomous and smart wireless network where the benefits of RISs are fully realized. 

Toward this goal, mushroom structures with varactor diodes have been used to achieve reconfigurable resonance, with incident waves coupled to substrate-integrated waveguides (SIWs) in the bottom layer through annular slots \cite{alamzadeh2021reconfigurable}. Although this design successfully integrated sensing and reflection in the same configuration, it required many RF chains for information extraction, significantly increasing system cost and complexity. A later refinement used only a few hybrid elements connected to SIWs to reduce complexity while enabling AoA detection through sparse sampling \cite{alamzadeh2022sensing}. The core concept underlying this HRIS was later experimentally demonstrated in a follow-up study \cite{alamzadeh2025experimental}. However, this device relied on multiplexing the incident wave across HRIS's elements before reaching the sensing waveguide. This led to weak signal strength, as the SIWs operated near cut-off due to space constraints. Furthermore, implementing SIWs and the required DC circuitry on the same substrate remains highly complex, making manufacturing previous HRIS designs challenging.


In this paper, we simplify the overall design of the HRIS by proposing  a 1-bit configuration with large element spacing and a parallel plate waveguide (PPWG) for collecting the sensed signal. Using 1-bit tuning with large element spacing ($~\lambda/2$), however, can cause unwanted quantization lobes that reduce efficiency and cause interference\cite{kashyap2020mitigating}. Increasing the number of tuning bits, e.g., by using varactor diodes, can mitigate this issue, but it increases complexity and cost. Instead, a more practical solution is phase randomization \cite{kashyap2020mitigating}, which involves introducing random phase delays in each unit cell. These random phase delays can disrupt the periodic patterns that cause quantization lobes. This method effectively suppresses the grating lobes while preserving the simplicity and efficiency of a 1-bit RIS.

Using elements spaced farther apart than in previous HRIS designs reduces element-to-element coupling. Earlier designs used this coupling to multiplex the signal incident on all elements into the one collected by a few hybrid meta-atoms. When the spacing between elements increases, the signal collected at the hybrid meta-atom is dominated by the signal incident directly on it. This can complicate the sensing process and degrade overall performance. To enhance the strength of the sensing signal collected across the entire RIS, we change the sensing geometry to a PPWG instead of using SIWs. We introduce small slots in the ground plane of each patch, allowing a portion of the signal to couple into the PPWG. This design ensures that some of the signal incidents on each element are coupled to the PPWG for sensing purposes. We use only two coaxial connectors to collect the data for sensing. These modifications from previous HRISs ensure a simple configuration that addresses both signal strength and the complexity of previous designs. To address quantization lobes, we implement phase randomization by randomly varying the size of the coupling slots along the HRIS.

Preliminary results confirming this novel design were presented at an earlier conference \cite{keshmiri2025novel}. In this paper, we expand upon that work to outline the design considerations for the proposed HRIS. We detail how randomized slot sizes are used to mitigate quantization lobes. We use full-wave simulations to demonstrate the proposed HRIS's ability to steer beams in desired directions without encountering quantization lobes. We also evaluate the system's performance in detecting the AoA using two methods: least squares solvers and multilayer perceptron (MLP). The accuracy in detecting AoA as a function of signal-to-noise ratio (SNR) is subsequently analyzed.

\section{METASURFACE ELEMENT DESIGN}
\begin{figure}[!t]
\centering
\includegraphics[width=\linewidth]{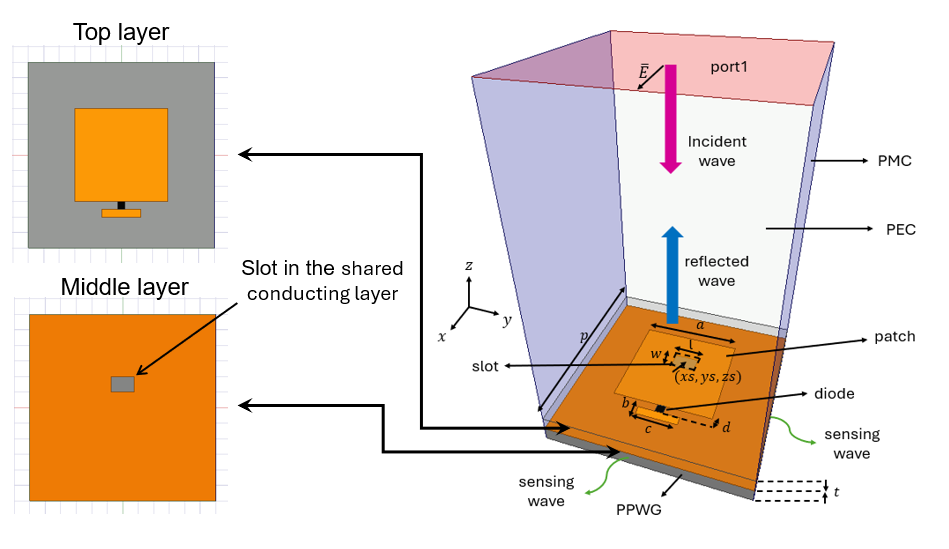}
\caption{Simulation configuration of the building block of the proposed HRIS. $(x_s, y_s, z_s) = (-3, 0, -1.624)$ which is a vector from the patch's center to the slot's center. Port 1 is de-embedded to the surfaces of the patch.}
\label{fig:fig1}
\end{figure}

\noindent The configuration of the building block for the proposed HRIS is illustrated in Fig.~\ref{fig:fig1}. It features a square patch element with a side length of 12 mm, implemented on a Rogers 4003 substrate with a thickness of 1.624 mm and a relative permittivity ($\epsilon_r$) of 3.55. This element is loaded with a PIN diode, which is modeled in its \textit{off} state as an effective capacitor of 0.15 pF and in its \textit{on} state as a resistor of 0.1 $\Omega$. The patch element shares a ground plane with a PPWG made from the same substrate material and thickness. A small rectangular slot is incorporated into the shared ground plane to allow a portion of the incident wave to couple into the PPWG. The design of this element was optimized for operation at a frequency of 5.6 GHz through simulations conducted in Ansys HFSS, and the dimensions are reported in Table~\ref{table:design_parameters}. While the design was specifically tailored for this frequency, it can be easily adapted to other frequencies by adjusting the geometry or the switchable component. As shown in Fig.~\ref{fig:fig1}, we placed the slot off center to reduce the coupling to the PPWG based on a parametric study in Ansys HFSS (not shown for brevity). A similar idea has been pursued \cite{alamzadeh2022sensing}. The slot length has a significant effect on performance and will be used later for phase control. The width of the hole, $w$, due to the polarization of the electric field, does not significantly impact the performance unless it becomes relatively large. Therefore, we selected it to be sub-wavelength to ensure that it does not adversely affect performance.

\begin{figure}[!t]
\centering
\includegraphics[width=\linewidth]{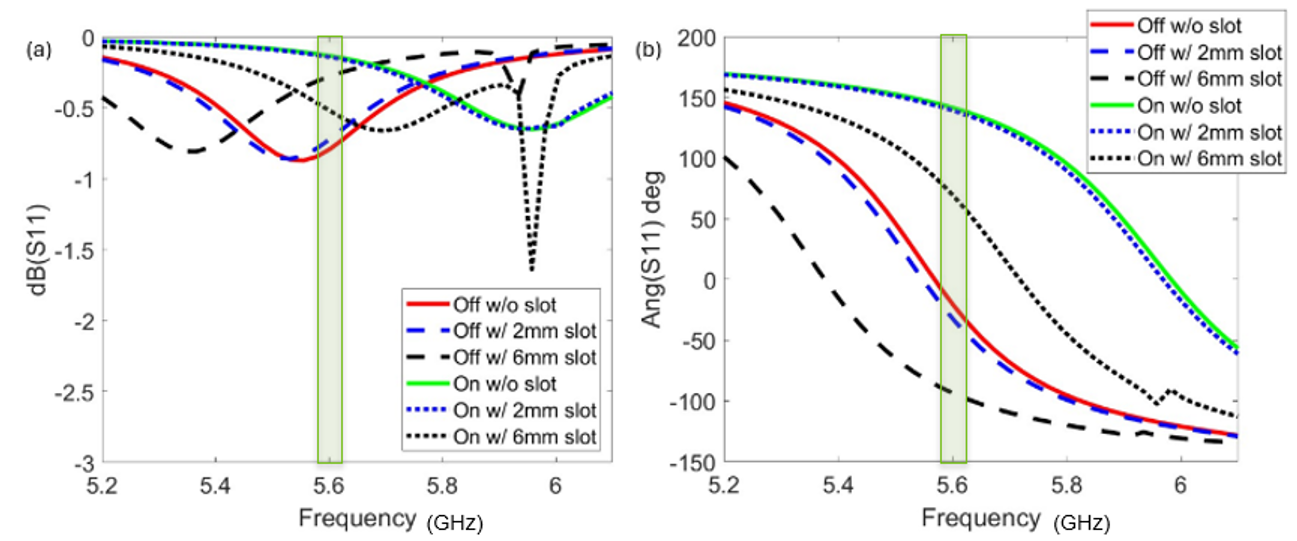}
\caption{(a) Amplitude and (b) de-embedded phase of the reflection coefficient of the proposed building block.}
\label{fig:fig2}
\end{figure}

The simulated reflection coefficient for the element, both with and without the slots, is depicted in Fig.~\ref{fig:fig2}. The results indicate that adding the slots reduces the reflection coefficient by less than 0.1 dB, suggesting that only a small portion of the signal is coupled to the PPWG. In these simulations, the slot was positioned at an offset of -3 mm from the center of the substrate in the x-direction. To achieve phase randomization, which is necessary for beamforming, we require two different slot lengths that produce four states with approximately $90^\circ$ phase differences relative to each other. As shown in Fig. 2(b), slot lengths of 2 mm and 6 mm meet this condition at a frequency of 5.6 GHz. 

\begin{table}[ht]
\centering
\caption{Design parameters for the metasurface element.}
\label{table:design_parameters}
    \begin{tabular}{|c|c|c|c|}
        \hline
        \textbf{Parameter} & \textbf{Value} & \textbf{Parameter} & \textbf{Value} \\
        \hline
        a & 12mm & b & 1mm \\
        \hline
        c & 5mm & d & 1mm \\
        \hline
        l & 2 or 6mm & p & 24mm \\
        \hline
        t & 1.624mm & w & 2mm \\
        \hline
    \end{tabular}
\end{table}

\section{BEAMFORMING}
\begin{figure}[!t]
\centering
\includegraphics[width=0.9\linewidth]{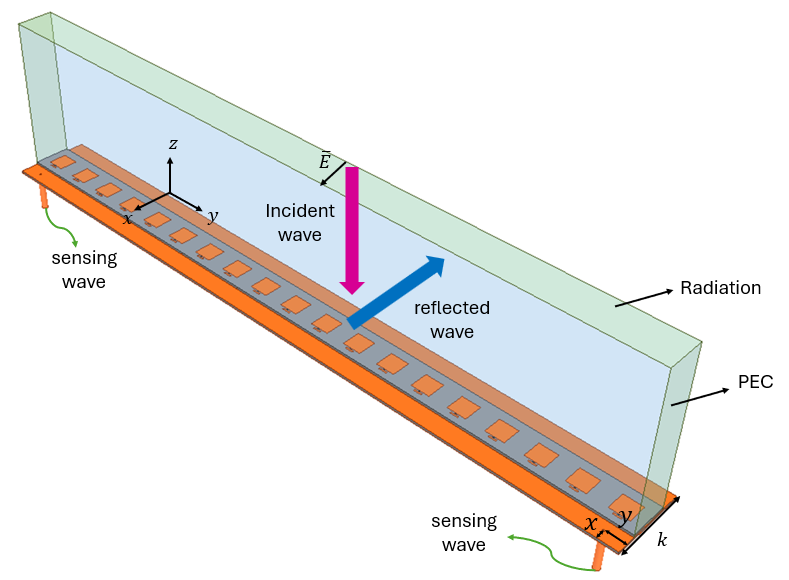}
\caption{Simulation configuration of an array of 19 elements made of the proposed building block. Here, $x=6mm , y=12mm , k=48mm.$}
\label{fig:fig3}
\end{figure}

\begin{figure}[!t]
\centering
\includegraphics[width=0.8\linewidth]{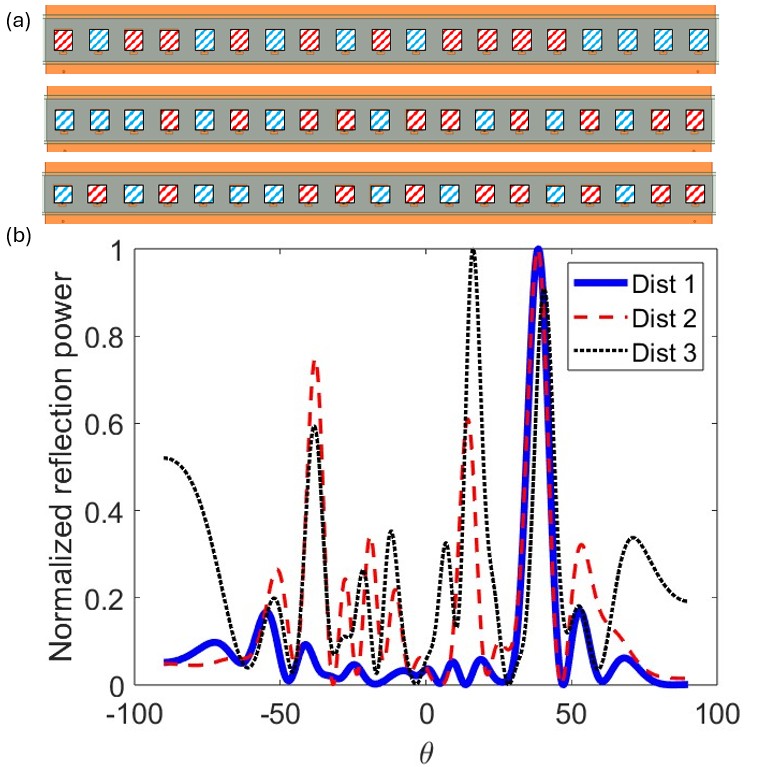}
\caption{ (a) 3 different layouts of the 1D array where elements with $2mm$ long slots and those with $6mm$ long slots are denoted by red and blue stripes, respectively. (b) Numerically calculated reflection pattern for different slot distributions depicted in (a).}
\label{fig:fig4}
\end{figure}

\noindent We analyze the beamforming capability of the proposed HRIS by simulating a 1D configuration consisting of 19 of the proposed elements (Fig.~\ref{fig:fig3}). To simplify the analysis, the beamforming and sensing capabilities are evaluated only in the yz-plane, while an infinite periodic array in the perpendicular direction is assumed. To add random phase delays to the elements, which is necessary to suppress the quantization lobe, each element is configured with a randomly assigned slot length of either 2 mm or 6 mm. Based on HFSS simulations, each slot length supports two distinct reflection phases depending on the on/off diode state (Fig.~\ref{fig:fig2}(b)). These discrete phase shifts were extracted and used in a simplified array-level simulation in MATLAB.
In these numerical studies, each element is modeled as a dipole whose phase will depend on the corresponding element's on/off status. We determine the on/off distribution such that the resulting phase gradient best approximates the desired progressive phase profile to steer the beam toward a target direction \cite{yu2011light}. We used this array calculation to estimate the far-field reflection pattern for different distributions of 19 elements with 2mm or 6mm slots. 
To find an effective configuration (with minimal grating lobes), we generated several random distributions of slot assignments and simulated the corresponding far-field patterns for different reflection angles. Fig.~\ref{fig:fig4} shows examples of this calculation, including the one ultimately selected based on its lowest side lobe levels. It is evident that the choice of the phase randomization can significantly impact the beam redirection performance. Thus, one can select an optimum distribution, an interesting task for future work.

\begin{figure}[!t]
\centering
\includegraphics[width=0.8\linewidth]{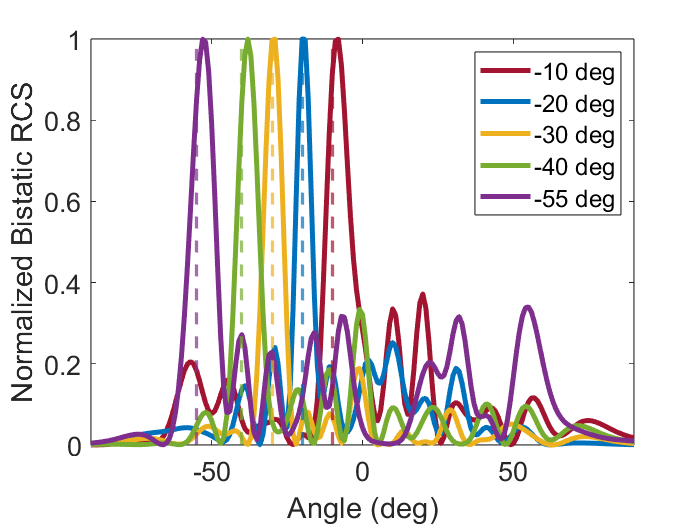}
\caption{Beamforming results for five different desired angles (denoted by dashed lines), assuming a normal incident wave.}
\label{fig:fig5}
\end{figure}

Once the distribution of slots for phase randomization is selected, we then design the distribution of on and off elements to form the desired reflection patterns. This is done by minimizing the Euclidean distance in the complex domain between the required phase for each element and the available reflection phase obtained from the element simulations (see  (Fig.~\ref{fig:fig2}(b)). Fig.~\ref{fig:fig5} illustrates the beamforming obtained in Ansys HFSS for various desired angles, showcasing the proposed HRIS's capability to steer beams with suppressed quantization lobes.

\section{Sensing}
\begin{figure}[!t]
\centering
\includegraphics[width=0.7\linewidth]{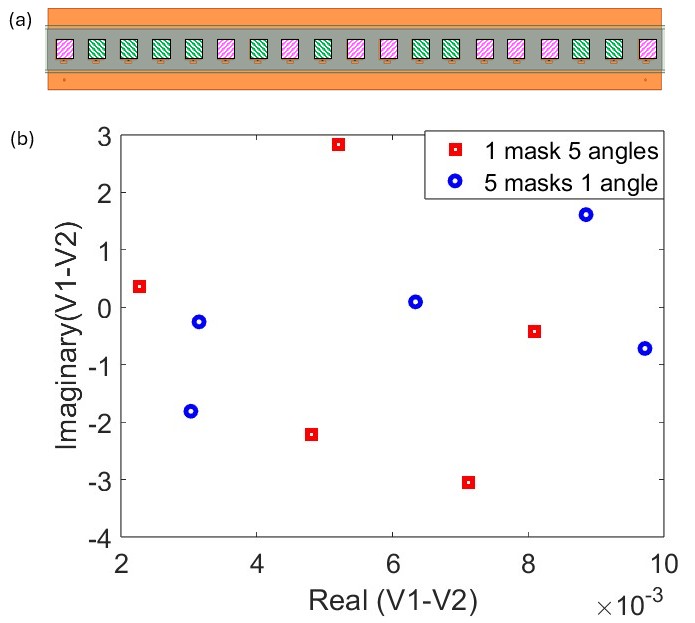}
\caption{(a) Example of on/off (respectively pink and green) distribution of elements. (b) Difference of the voltages induced on the coaxial connectors due to incident signals with five different incident angles for the mask shown in (a) and for the case of 1 incident angle and 5 different random masks.}
\label{fig:fig6}
\end{figure}

\noindent Next, we examine the sensing capabilities of the proposed HRIS. To minimize complexity, we implement AoA detection using only two coaxial connectors to collect the sensing signal. It is worth noting that the signal collected at the coaxial connectors is the multiplexed version of the signals coupled from all elements into the PPWG. 
Figure \ref{fig:fig6} illustrates the differences in voltages induced on the coaxial connectors as the AoA changes. This observation confirms that the signal measured at the coaxial connectors contains information regarding the AoA. However, this relationship is not trivial and cannot be easily inverted. Additionally, more measurements are required to accurately determine the AoA. To address this issue, we utilize random distribution of on/off elements (called masks for brevity). As shown in Figure \ref{fig:fig6}, applying different masks for each AoA yields new measurements, improving the condition number of the inverse problem.

\begin{figure}[!t]
\centering
\includegraphics[width=\linewidth]{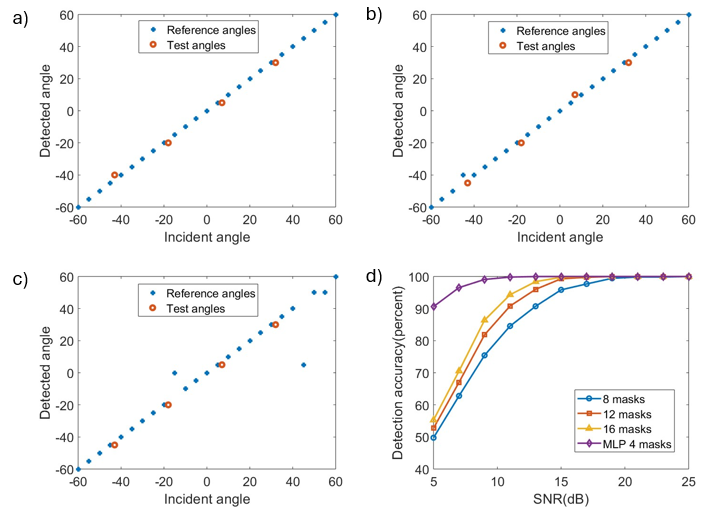}
\caption{Detected versus the actual AoA for the case with (a) 16 masks and SNR=20 dB, (b) 16 masks and SNR=15 dB, (c) 8 masks and SNR=15 dB (d) detection accuracy for different SNR levels and number of masks. }
\label{fig:fig7}
\end{figure}

To detect AoA, we followed a process similar to that previously reported ~\cite{alamzadeh2022sensing}. We first discretized the possible AoAs into $N$ reference angles, ranging from $-60^\circ$ to $60^\circ$ in $5^\circ$ increments, resulting in 25 reference angles. The sensing matrix, $\mathbf{H}_{M \times N}$, is then constructed by collecting the difference between the voltages on the coaxial connectors caused by incident waves arriving from reference angles using $M$ masks with random on/off distribution. The estimation problem then follows the formulation:  

\[
\mathbf{g}_{M \times 1} = \mathbf{H}_{M \times N} \mathbf{f}_{N \times 1},
\]

\noindent where $\mathbf{g}$ represents the measurements for an arbitrary incident signal. The ith element of $\mathbf{f}$ is zero if the incident AoA coincides with the ith reference angle and 0 otherwise. Since $\mathbf{H}$ is not square, we used a computational technique (i.e. Conjugate Gradient Squared (CGS)) to estimate $\mathbf{f}$. The resulting estimated vector, $\mathbf{f_{est}}$, will not be all 0s and 1s. To deduce the AoA, we consider the index of maximum of $|\mathbf{f_{est}}|$ to be the estimated AoA. If the incident signal does not align with a reference angle, the closest reference angle is used for the estimation. Fig.~\ref{fig:fig7}(a-c) presents the detected angle versus the actual incident angle, including reference angles used to form the sensing matrix and additional test angles not included in the reference set for different SNR levels and different mask numbers. The results demonstrate that the proposed HRIS system can estimate AoA with an accuracy of $\pm 3^\circ$ (when $M=16$ and SNR=20 dB, shown in Fig.~\ref{fig:fig7}(a)). In Fig.~\ref{fig:fig7}(b), the SNR level is reduced to 15 dB. A slight degradation in estimation accuracy can be observed. Fig.~\ref{fig:fig7}(c) shows the results with only 8 masks at an SNR of 15 dB. In this case, the estimation performance noticeably decreases, which shows that high detection accuracy at low SNR levels will require more masks.

We conducted Monte Carlo simulations to evaluate the detection accuracy for different scenarios more systematically. For each SNR value, 50 independent trials with new instances of the added noise were performed. In each trial, the AoA was estimated using the proposed sensing method. The detection was considered successful if the estimated angle was within $\pm 3^\circ$ of the true incident angle. The average detection accuracy across the 50 trials was calculated to generate each data point in Fig.~\ref{fig:fig7}(d). This process was repeated for configurations with 8, 12, and 16 masks. Fig.~\ref{fig:fig7} (d) shows how the detection accuracy increases with higher SNR levels across all configurations, and a larger number of masks (12 or 16) yields better performance (than 8 masks) at low to moderate SNRs; however, the effect of mask number diminishes as the SNR exceeds 20\,dB, where all configurations approach nearly 100\% accuracy.

To overcome the sensitivity of CGS to reduced measurements and noise and to enable accurate detection with fewer spatial masks, we introduce a neural network-based approach using a multilayer perceptron (MLP) classifier. The same set of 25 incident angles used for CGS was used to construct the training data, where each sample consisted of 16 features representing the real and imaginary parts of the voltages measured across four distinct masks. The MLP was trained on synthetically expanded data, generated by repeating these 25 samples 300 times and adding Gaussian noise corresponding to an SNR of 50\,dB to all entries to improve generalization.

The model architecture included three hidden layers with ReLU activations, batch normalization, and dropout regularization. The model was evaluated on the test portion of the same dataset, with additional Gaussian noise added to simulate varying SNR levels from 5\,dB to 25\,dB in 2\,dB increments. Using the same Monte Carlo procedure as CGS, the classification was repeated 50 times for each SNR value with independently generated noise samples added to the test data. Detection accuracy was calculated based on whether the predicted class was within $\pm$1 bin (i.e., $\pm$3$^\circ$) of the true angle class. As shown in Fig.~\ref{fig:fig7}(d), the MLP classifier demonstrates a strong ability to maintain accurate angle classification across a wide range of SNRs, even when trained using only measurements from four masks.

\section{CONCLUSION}
\noindent This paper presented a novel 1-bit HRIS that integrates AoA detection and beamforming capabilities while maintaining a simple and cost-effective design. The proposed HRIS leverages phase randomization to mitigate quantization lobes, improving beamforming efficiency despite using only 1-bit tunability and large element spacing. The sensing mechanism relies on randomly structured slots that couple a portion of the incident wave into a PPWG, where two coaxial connectors collect the signal. By applying compressive sensing techniques as well as MLP, the system effectively estimates AoA with an accuracy of ±3°, even under low SNR conditions. Simulation results demonstrate that the proposed HRIS achieves robust beamforming performance while simultaneously detecting the propagation environment without requiring additional feedback loops. This low-complexity and power-efficient architecture makes the HRIS a promising candidate for applications in smart wireless communication, intelligent wireless power transfer configurations, and RF sensing systems.

\section{ACKNOWLEDGMENT}
\noindent This material is based upon work supported by the National Science Foundation under Grant No. ECCS-2333023.

\bibliographystyle{ieeetr}
\bibliography{reference}

\end{document}